\documentstyle[12pt, epsf,epsfig,a4wide]{article}

\begin{document}
\begin{titlepage}

\begin{flushright}
hep-th/0205298
\end{flushright}

\begin{centering}
\vspace{0.1in}

{\Large {\bf Vacuum Energy  and Cosmological Supersymmetry Breaking
in Brane Worlds}}

\vspace{0.2in}

{\bf Elias Gravanis and Nick E. Mavromatos } \\
\vspace{0.05in}
Department of Physics, Theoretical Physics, King's College London,\\
Strand, London WC2R 2LS, United Kingdom.

\vspace{0.4in}
 {\bf Abstract}

\end{centering}

\vspace{0.2in}

{\small 
In the context of a toy model we discuss 
the phenomenon of colliding five-branes, 
with two of the extra 
space dimensions compacified on tori. In one of the branes (hidden world)
the torus is magnetised. Assuming opposite-tension 
branes, we argue that the collision 
results eventually in 
a (time-dependent) cosmological vacuum energy,  whose value today 
is tiny, lying comfortably
within the standard bounds by setting the breaking
of the four-dimensional supersymmetry at a TeV scale. 
The small value of the vacuum energy as compared with the 
supersymmetry-breaking scale is attributed to 
transient phenomena with relaxation times 
of order of the Age of the Universe.
An interesting feature of the approach is the absence 
of a cosmic horizon, thereby allowing for 
a proper definition of an $S$-matrix. 
As a result of the string non-criticality
induced at the collision,  
our model does not provide 
an alternative to inflation, given that arguments can be given 
supporting 
the occurence of an inflationary phase
early after the collision. The physics before the collision
is not relevant to our arguments on the cosmological constant
hierarchy, which are valid for asymptotically long times after it.}

\end{titlepage}

One of the most  important unsolved puzzles 
in Theoretical Particle Physics
is the issue of the smallness of the Cosmological Constant
(or, better vacuum energy density) in comparison with other
physical scales, for instance, 
the scale at which Supersymmetry is broken
in supersymmetric theories. 
The resolution of such puzzles may lie in the 
way by which supersymmetry is broken. One  
interesting idea is that supersymmetry is broken somehow
cosmologically, in the sense of its breaking being linked
to a non-zero cosmological constant. Such an idea
has been studied recently in~\cite{banks} in the modern context
of brane (M) theory.

In ref. \cite{banks}, the vacuum energy has been assumed constant.
This might not be necessarily the case, though. One might 
encounter transient situations, as in 
quintessence models~\cite{carroll}, where
the ``vacuum'' energy is relaxing to zero asymptotically
by some power (usually quadratic) of the Cosmological-frame time.
Such scenaria are interesting, since they allow for an eventual exit from 
a de Sitter phase, implying non-eternally accelerating 
Universes. This is a welcome fact from the point of view 
of string-theory~\cite{dgmpp},
given that eternally accelerating (de Sitter) Universes
have cosmic horizons, which makes a definition  
of a 
S(cattering)-matrix connecting asymptotic states
problematic~\cite{eternal,emnsmatrix}.

In this article we shall adopt this latter point of view,
and present a scenario, albeit crude, according to which
colliding branes in superstring theory may result in
a way of breaking supersymmetry on our four-dimensional
world at a TeV scale, while maintaining a very small
vacuum energy, decreasing with cosmological time. 
We should notice that scenaria with time-dependent 
vacuum energies have been considered by many authors
in the past~\cite{timedep}. 
However, the physics of our model as well as its focus 
are different. We shall be interested in attempting to resolve
the issue of the hierarchy between the 
cosmological vacuum energy and the supersymmetry-breaking scale.  
The relaxation rate of the vacuum energy 
is found to be proportional to $1/t^2$, where $t$ is the 
Robertson-Walker time. This is argued to be sufficient 
for a resolution of the 
cosmic horizon problem as well.

To commence our discussion, 
let us consider for definiteness two five-branes of type IIB strings, 
embedded in a ten dimensional bulk space time. 
Two of the longitudinal brane dimensions are assumed compactified
on a small torus, of 
radius $R$. In one of the branes,
from now on called {\it hidden}, the torus is {\it magnetized} 
with a 
constant magnetic
field of intensity H. This amounts to an effective 
four-dimensional vacuum energy in that brane of order:
$V_{\rm hidd} = R^2H^2 > 0$. Notice that such compactifications
provide alternative ways of breaking supersymmetry~\cite{bachas},
which we shall make use of in the current article.

In scenaria with two branes embedded 
in higher-dimensional bulk space times, e.g. in the scenario
of \cite{randal}, it is natural to assume (from the point of view
of solutions to bulk field equations) that the two branes
have {\it opposite} tensions. We, therefore, assume that 
before the collision the visible 
brane (our world) has negative tension $V_{\rm vis}=-V_{\rm hidd} < 0$.
A negative tension brane is consistent with the possibility of 
accepting supersymmetric theories on it (anti-de-Sitter type).

The presence of opposite tension branes implies that 
the system is not stable. For our purposes we assume that the two 
branes are originally on collision course
in the bulk, with a relative velocity $u$. 
The collision takes place at a given time moment. 
This constitutes an event, which 
in our scenario is identified with the {\it initial
cosmological singularity} (big bang) on the
observable world. We note that 
similar scenaria
exist in the so-called ekpyrotic model for the Universe~\cite{ekpyrotic}.
It must be stressed, though, that 
the similarity pertains only to the brane-collision event. 
In our approach the physics is entirely different from 
the ekpyrotic scenario. First of all, the collision  
is viewed as an event resulting in non-criticality 
(departure from conformal invariance) of the underlying string theory,
and hence in non-vanishing $\beta$ functions at a $\sigma$-model level. 
On the contrary, in the scenario of \cite{ekpyrotic} the 
underlying four-dimensional effective theory (obtained after integration
of the bulk extra dimensions~\cite{ekpyrotic,linde}) is assumed always
critical, satisfying classical equations of motion, and hence vanishing 
$\sigma$-model $\beta$
functions.
Indeed this latter property leads only to contracting 
and {\it not expanding} four-dimensional
Universes according to the work of \cite{linde}, which constitutes
one of the main criticisms of the ekpyrotic universe. 

On the other hand, in our non-critical description of the collision
we do not assume classical solutions of the equations of motion,
neither specific potentials associated with bulk branes, as 
in \cite{ekpyrotic}. 
In our approach, we are interested only in the period 
{\it after the collision}. In fact, as we shall discuss in this article, 
in order to be able to use  
$\sigma$-model perturbation theory, one must restrict oneself
at times much longer after the collision. Before the collision
the moving branes may indeed be viewed as solutions of some 
classical equations,
as in \cite{ekpyrotic,linde}. But the collision-induced deviations from 
conformal invariance, we advocate here, play a r\^ole analogous to a 
sort of (stringy) phase transition. In our work we shall only be 
interested in the phase after the collision, where the degrees of 
freedom of the system and its description may be different:
the presence of non-criticality necessitates the introduction
of a whole new target-space dimension, the Liouville mode~\cite{ddk}. 

Notably, our perturbative approach is not valid for times near the collision,
where the string theory is strongly coupled, in contrast to the 
ekpyrotic Universe case. Moreover, the deviation from conformal 
invariance, quantified through the appearance of a central charge deficit
$Q^2$, which depends itself on time, is responsible for the entirely different
way of obtaining the fate of the four-dimensional cosmology in our case.
As we shall see, 
the generalized conformal invariance conditions (\ref{liouvcond}), 
stemming from the 
Liouville dressing of the non-critical theory~\cite{ddk}, 
encode the full dynamics of the 
four-dimensional theory in our approach. 
This dynamics, upon the identification of the Liouville mode with 
time, in a sense that will be specified in our article below, 
leads to asymptotically expanding Universes, in contrast to the 
contracting Universe situation of the ekpyrotic scenario~\cite{linde}. 

Most importantly, it must be stressed that 
our toy model should by no means be viewed
as an alternative to inflation, as claimed to be the case of 
the ekpyrotic Universe~\cite{ekpyrotic}.
In fact this point appeared to be the main focus of criticism of that
scenario~\cite{linde}. In our case, inflationary phases of the Universe
do exist, as demonstrated recently in the context of non-supersymmetric
type 0 strings upon deviations from criticality by either quantum 
fluctuations or brane 
collisions~\cite{dgmpp}. 
The presence of a time-dependent central-charge deficit
$Q^2(t)$ is crucial to the effect. As argued in \cite{dgmpp}, 
the effective four-dimensional theory, obtained after appropriate 
compactification
or integration over bulk dimensions, has at early times a phase
where inflation - at least in the sense of 
exponential expansion of the scale factor - 
always occur, succeeded by graceful exit from this 
de-Sitter-type phase, which is not possible in critical strings.

In our article we shall not be interested in such early 
times or such important issues as
density fluctuations {\it etc.}. Our toy model is too crude to 
allow for a full study of a present-day cosmology with matter. 
Instead we would like 
to make an interesting observation, by means of this toy model, 
according to which the abovementioned
string non-criticality leads, for 
asymptotically long times after the collision (including present eras), 
to a natural explanation of a hierarchy between the vacuum 
energy 
and the scale of supersymmetry breaking, as well as 
the lack of a cosmic horizon.
Nevertheless, we stress again, this toy non-critical string 
model is expected,
on the basis of the work of \cite{dgmpp}, to exhibit 
an inflationary phase and eventual graceful exit from it.

We now make the plausible assumption that, during the collision, 
there is electric current transfer 
from the hidden to the visible brane, which results in 
the appearance of a magnetic field on the 
visible brane. We also assume that the entire effect
is happening very
slowly and amounts to a slow flow of energy and current
density from the positive energy density brane to
the one with negative tension. In turn, this results in a
positive energy component
of order $H^2R^2$  
in the vacuum energy of the visible brane world.
This energy component may be 
assumed to {\it cancel } the pre-existing negative tension asymptotically
in time, leading to a vanishing cosmological constant at $t=\infty$. It is 
our aim to find, by a preliminary $\sigma$-model analysis,
the asymptotic form (in large times) of this time-dependent 
four-dimensional vacuum energy, and  
relate this to supersymmetry breaking.
Notice that such a scenario 
imitates a slow relaxation period of the Universe, which still goes on.
This is in accordance with quintessence models~\cite{carroll}
which have not yet reached their equilibrium state.

We should notice at this stage that the initial instability 
due to the negative tension brane disappears from the 
observable sector, given that the cosmological time flow
begins from the moment of the collision. 
As we shall discuss in some detail in this article, 
at the moment of the collision the conformal invariance 
of the $\sigma$-model describing excitations on the observable
world is spoiled, thereby implying the need for Liouville 
dressing~\cite{ddk,emn}. This procedure restores 
conformal invariance at the cost of introducing 
an extra target space coordinate (the Liouville mode $\phi$),
which in our model has time-like signature.
Hence, initially, one
faces a two-times situation. 
We argue, though, that
our observable (cosmological) time $X^0$ parametrizes a certain curve, 
$\phi={\rm const.}\;X_0+{\rm const.'}$, on the
two-times plane $(X_0,\phi)$, and hence one is left with one
physical time.

The appearance of the magnetic field on the visible 
brane, on the dimensions $X^{4,5}$, 
is described (for times long after the collision)
within a $\sigma$-model
superstring formalism by the boundary deformation~\cite{abouel}:
\begin{equation} 
{\cal V}_{H}= \int _{\partial \Sigma} 
A_5\partial_\tau X^5 -i F_{05} \overline{\psi}^0 \rho^0 {\psi}^5 -i F_{45} \overline{\psi}^4 \rho^0 {\psi}^5 
\label{magfield}
\end{equation} 
where $A_5=e^{\varepsilon X^0} H X^4$ and $F_{\mu\nu}$ is the (abelian) field strength of
$A_{\mu}$, $X^0$ is the time and
$\partial_\tau$ denotes tangential $\sigma$-model derivative
on the world-sheet boundary.
The $\sigma$-model 
deformation (\ref{magfield}) 
describes open-string excitations attached to the brane world.
In our approach, 
for convenience, we have set 
the charges at the end of the open string on the visible world
equal to one. In (\ref{magfield}) the presence of the quantity 
$\varepsilon \to 0^+$
reflects the {\it adiabatic} switching 
on of the magnetic field
after the collision. It should be remarked that in our approach 
the quantity $\varepsilon$ is viewed as a 
renormalization-group scale parameter, which, as we shall argue 
below, flows in such a way 
that any contribution from 
the exponent $\varepsilon X^0$ to $H$ is cancelled
after Liouville dressing~\footnote{At this point we should
remark that one could have used
a different way of parametrizing the 
adiabtaic switching on of the magnetic field, for instance
a function $H(1 - e^{-\varepsilon X^0})$, $\varepsilon \to 0^+$.
The conformal field theory analysis in that case is 
similar to the case considered above, and will not be presented
here.}.  

In addition to the magnetic field deformation, the $\sigma$-model 
contains also boundary deformations describing the `recoil' of the 
visible world due to the collision:
\begin{equation} 
{\cal V}_{\rm rec} = \int _{\partial \Sigma}  Y_6(X_0)\partial _n X^6+ i \partial_0 Y_6 \; \overline{\psi}^0 \rho^1 {\psi}^6
\label{recoil}
\end{equation}
where $Y_6(X_0)=u X^0 e^{\varepsilon X^0}$,
$\partial_n$ denotes normal $\sigma$-model derivative
on the world-sheet boundary and we have assumed for simplicity that
the motion of the branes is along the sixth bulk dimension. In (\ref{recoil})
$u$ denotes the recoil velocity of the visible world, which is 
of the order of the 
incident velocity of the hidden brane~\footnote{Notice that
a similar formalism describes also a plastic collision, where
the two branes merge to a single one after the collision.}. 

As can be seen straigthforwardly, by an operator-product-expansion 
analysis with the free string world-sheet stress tensor,  
the presence of the exponential 
$e^{\varepsilon X^0}$ implies a small but negative 
world-sheet anomalous dimension
$-\frac{\varepsilon ^2}{2} < 0$, and hence the relevance of 
both operators (\ref{magfield}),(\ref{recoil}) from a renormalization-group
point of view. By virtue of the Zamolodchikov's c-theorem~\cite{zam} there is 
a central charge deficit $Q^2$, whose rate of change with the 
renormalization-group scale on the world sheet ${\cal T}$ is:
\begin{equation} 
\frac{d}{d{\cal T}}Q^2 = -\beta^i {\cal G}_{ij} \beta^j 
\label{cthe}
\end{equation}
A straightforward computation of the two point 
correlators between the operators ${\cal V}_H, {\cal V}_{\rm rec}$
yields the Zamolodchikov metric in coupling constant space~\cite{zam}:
\begin{equation} 
{\cal G}_{HH} = |z|^4\langle e^{\varepsilon X^0 (z)}X^4(z)\partial_\tau X^5
(z)e^{\varepsilon X^0 (0)}X^4(0)\partial_\tau X^5 (0)\rangle \sim 
e^{4\varepsilon^2 {\rm ln}|L/a|^2} {\rm ln}|L/a|^2 
\end{equation}
and similarly for ${\cal G}_{uu}$. The non-diagonal elements
of ${\cal G}_{ij}$ vanish. 
It can be easily checked that  
the contributions from the world-sheet fermionic fields are subdominant 
as compared with the 
bosonic ones.
We may identify 
\begin{equation}
\varepsilon ^{-2} \sim {\rm ln}|L/a|^2
\label{epsilscale}
\end{equation} 
so that the above correlators scale as spacetime length squared. 
It is a rather established fact that such 
an identification is natural, if not unavoidable, once one introduces 
simple operators with
anomalous dimension related to a new spacetime scale in their 
definition~\cite{kogan,frw}.
The above considerations  imply that the Zamolodchikov metric is 
singular in the limit $\varepsilon \to 0^+$:
\begin{equation}
{\cal G}_{HH} \sim {\cal G}_{uu} \sim \frac{1}{\varepsilon ^2} 
\label{metriczam}
\end{equation}

On the other hand, the 
$\sigma$-model $\beta$-functions for the couplings $H$ and $u$, 
corresponding
to the vertex operators (\ref{magfield}) and (\ref{recoil}) 
respectively,
are:
 $\beta_{\bar H} = \frac{d {\bar H}}{d {\cal T}}= 
-\frac{\varepsilon^2}{2} {\bar H} $, 
$\beta_{\bar u} = \frac{d {\bar u}}{d {\cal T}}= 
-\frac{\varepsilon^2}{2} {\bar u} $,
where the barred notation pertains to renormalized (scale $\varepsilon$ 
dependent) quantities, and ${\cal T} \propto 
\varepsilon^{-2} \sim {\rm ln}|L/a|^2$ from (\ref{epsilscale}). These 
relations imply that 
the scale-$\varepsilon$  dependent couplings have the form~\cite{szabo}: 
$ {\bar H} \equiv \varepsilon H$,
$ {\bar u} \equiv \varepsilon u$, where $H,u$ are scale-$\varepsilon$ 
independent quantities.

From the above considerations one arrives at the following 
differential equation for the central charge deficit:
\begin{eqnarray}
&~& \frac{d}{d {\cal T}}Q^2 \sim - \frac{H^2 + u^2}{{\cal T}^2} \quad \to
\quad 
Q^2 ({\cal T}) = Q_0^2 + \frac{H^2 + u^2}{{\cal T}}
\label{ratedeficit}
\end{eqnarray}
where, for formal completeness we give here the more general
case of $n$ compactified tori, with  
$n=1$ corresponding to a five brane, $n=2$ to a seven brane
and $n=3$ to a nine brane, which exhausts the possibilities
in the case of type IIB superstring we are dealing for 
definiteness here. 
The quantity 
$Q_0^2 = Q^2 (\infty)$ is a constant, and 
consists of the vacuum energy density contributions
of the visible brane world $V_{\rm vis} <0 $ and the energy density of the 
magnetic field $H^2R^{2n} >0$.
A physical meaning to (\ref{ratedeficit}) 
can be given by noting that its $H$-dependent term represents 
the electric-field energy density on the brane  
$\int_{\rm tori} F_{05}^2 \propto \varepsilon^2 H^2 R^{2n}$, 
induced by the time-varying magnetic 
field $He^{\varepsilon X^0}$. The $u$-dependent term on the other hand
represents recoil kinetic energy contributions, which in our case
are subleading. 
As we shall explain later on, the relaxation situation we  
encounter here implies that $Q_0$ is the equilibrium
vacuum energy density, which we take 
to be zero $Q_0^2 =0$ due to the cancellation between the initial
vacuum energies of the colliding branes~\cite{randal}.

The non-conformal deformed $\sigma$-model can become conformal
as usual by Liouville dressing~\cite{ddk}. Let $\varphi$ be 
the Liouville mode with $\sigma$-model action 
\begin{equation} \label{liouvact}
{\cal L}_\phi = \int _\Sigma 
Q^2({\cal T}) \partial \varphi  {\overline \partial \varphi}
+ \int _\Sigma R^{(2)} Q^2({\cal T}) \varphi + \dots 
\end{equation}
where the zero mode of $\varphi $ is related to the renormalization-group 
scale ${\rm ln}|L/a|^2 \sim {\cal T}$, 
being viewed as a covariant renormalization
scale on the world sheet~\cite{emn}. The $\dots $ in (\ref{liouvact}) 
express
possible world-sheet 
boundary extrinsic  curvature terms, with which we shall not deal
explicitly here.

It is customary~\cite{ddk} to normalise the kinetic term of the 
Liouville action by rescaling 
\begin{equation} 
\varphi \to \phi \equiv Q({\cal T})\varphi~,
\label{rescaling}
\end{equation} 
which plays the role of an extra target space-time dimension.
Due to the supercritical nature
of the central charge deficit (\ref{ratedeficit})
at scales ${\cal T} < \infty$, $Q^2({\cal T}) > 0$, 
the extra Liouville dimension is {\it time like}~\cite{aben}, 
and therefore one
faces a two-target-times situation. 

We now come to discuss the dressing of the vertex operators  ${\cal V}_H, {\cal V}_u$.
This amounts to introducing the Liouville field in the definitions of $A_5,Y_6$, that
operates as one more spacetime coordinate resulting in conformally
invariant boundary deformations. The new fields are given by
\begin{equation}
A_5(X_0,X_4,\phi)=H X^4e^{\varepsilon X^0+ \alpha \phi} , \;\; 
Y_6(X_0,\phi)=u X^0 e^{\varepsilon X^0+ \alpha \phi}
\end{equation}
The world-sheet supersymmetrized vertex operators are now given by
\begin{eqnarray}
{\cal V}_{H}&=& \int _{\partial \Sigma} 
A_5\partial_\tau X^5 -i F_{05} \overline{\psi}^0 \rho^0 {\psi}^5 -i F_{45} \overline{\psi}^4 \rho^0 {\psi}^5 
-i F_{\phi5} \overline{\psi}^{\phi} \rho^0 {\psi}^5\nonumber \\
{\cal V}_{\rm rec} &=& \int _{\partial \Sigma}  Y_6\partial _n X^6+ i \partial_0 Y_6 \; 
\overline{\psi}^0 \rho^1 {\psi}^6 +i \partial_{\phi} Y_6 \;\overline{\psi}^{\phi}\rho^1 {\psi}^6 
\label{dressed}
\end{eqnarray}
where ${\psi}^{\phi}$ is the supersymmetric partner of the Liouville field. 
The gravitational (Liouville) anomalous dimensions $\alpha$ 
are given by~\cite{ddk}:
\begin{equation}\label{gravad} 
\alpha = -\frac{Q({\cal T})}{2} + \sqrt{\frac{Q^2({\cal T})}{4} + 
\frac{\varepsilon^2}{2}} 
\end{equation} 
As a $\sigma$-model, this $(d,2)$ theory is conformal~\cite{ddk}. 

From the work of \cite{bachas} it becomes clear that the coupling constant 
$H$ is associated with supersymmetry-breaking mass splittings.
This has to do with the different way fermions and bosons couple to an 
external magnetic field. The mass splittings squared of an open string 
are generically of order $\delta m^2 \sim H$.
To be precise, in the case of a constant magnetic field, examined in 
\cite{bachas}, the supersymmetric mass splittings are
\begin{equation} 
\Delta m_{\rm string}^2 = 4 H \Sigma_{45}
\end{equation}
with $\Sigma_{45}$ the spin operator on the plane of the torus.
To ensure the phenomenologically reasonable order of magnitude of a TeV scale,
one must assume very small~\cite{bachas} $H \sim 10^{-30} \ll 1$ 
in Planck units.
In a similar manner, one assumes naturally that the velocities $u$ are 
also much smaller than one, in order for our perturbative world-sheet
analysis to be valid~\cite{kogan,szabo}. 
For such small values of the couplings $H,u$ one has 
from (\ref{ratedeficit}), (\ref{gravad}) that 
the Liouville anomalous dimensions are of order 
$\alpha \sim \frac{\varepsilon}{\sqrt{2}} $, ignoring 
H,u dependent terms, which are subleading for $\varepsilon \ll 1$.  

In our case, we have a slowly varying magnetic field $He^{\alpha \phi + 
\varepsilon X^0}$, from which we may deduce approximately the corresponding
mass squared splittings:
\begin{equation} 
\Delta m_{\rm string}^2 \sim   He^{\alpha \phi + 
\varepsilon X^0 }\Sigma_{45} 
\label{split}
\end{equation}
The so-obtained mass splittings are constant upon the requirement 
that the flow of time $X^0$ and of Liouville mode $\phi$ are correlated
in such a way that 
\begin{equation} 
\varepsilon X^0 + \varepsilon \phi/\sqrt{2} 
={\rm constant}~,
\label{liouvtime}
\end{equation}
or at most slowly varying.
Notice that
deviations from the condition (\ref{liouvtime}) 
would result in very
large negative-mass squares, which are clearly unstable configurations.
Hence, the identification (\ref{liouvtime}) seems to provide
a resolution of this problem~\footnote{Note, however, that 
if one used the alternative representation of the adiabatic
switching on of the magnetic field $H(1-e^{-\varepsilon X^0})~, \varepsilon > 0$, the masses would be finite as $X^0 \to \infty$. Nevertheless
the condition (\ref{liouvtime}) would still be necessary from the 
physical point of view of having a single observable temporal coordinate
in space time. We discuss some consequences of the case where $\phi$ and 
$X^0$ are independent variables later on in the article.}.

The condition (\ref{liouvtime}) 
implies a connection of the zero mode of the Liouville field, ${\cal T}$, 
with the target time $X^0$. In this sense $Q_0^2 = Q^2 (\infty)$ represents
the central charge deficit of the theory asymptotically in time.
Given that the initial vacuum energy on the observable brane 
is assumed to be cancelled during the collision with the hidden world,
where the flow of our cosmic time (and hence the Liouville scale) starts,
it is natural to assume that $Q_0^2 =0$, which justifies our choice above. 
Note also that parametrizing this condition as $X^0=t$, $\phi_0 =\sqrt{2}t$,
and taking into account that, for convergence of $\sigma$-model 
path integration, it is formally necessary to work with Euclidean signature 
$X^0$~\cite{kogan}, the induced 
metric on the hypersurface (\ref{liouvtime}) in the 
extended space time  
acquires a 
Minkowskian-signature Robertson-Walker form: 
\begin{equation} 
ds^2_{\rm hypersurf} = -(d\phi_0)^2 + (dX^0)^2 + \dots = - (dt)^2 + \dots~.
\end{equation} 
where $\dots$ denote spatial parts. 

At this stage an important comment is in order regarding 
the stability of the condition (\ref{liouvtime}) in the 
context of Liouville strings. 
In our approach so far we have assumed a situation in which the 
magnetic field is adiabatically switched on after the collision 
and then asymptotes to a constant value. 
On the other hand, one may consider an equally plausible 
situation in which the magnetic field is switched on 
on our world, due to transient phenomena described above, 
and then 
relaxes to zero again. In such a case the magnetic field 
intensity on the brane world assumes  
the form 
\begin{equation} 
H\left(\Theta_\varepsilon (-X^0) + \Theta_\varepsilon (X^0)\right)
\label{twothetas}
\end{equation} 
where $\Theta_\varepsilon (X) = -i\int \frac{d\omega}
{\omega -i\varepsilon}e^{i\omega X}$, $\varepsilon 
\to 0^+$, denotes 
the regularized Heaviside function~\cite{kogan}.
A contour integral representation yields 
$\Theta_\varepsilon (X) = \theta (X) e^{-\varepsilon X}$, 
with $X > 0$ and  
$\theta (X)$ the conventional Heaviside (unregularised) function.

In this case, one obtains {\it a pair} of independent 
$\sigma$-model deformations,
corresponding to the two $\Theta$ functions in (\ref{twothetas}).  
The Liouville dressing procedure is now a bit more
complicated, but as we shall argue below, this case yields indeed
the dynamical stability requirements for the condition (\ref{liouvtime}). 
To this end, we
first remind the reader 
that in our previous analysis we have used 
the Liouville anomalous dimensions (\ref{gravad}). 
The restoration of conformal
invariance by Liouville dressing, however, actually 
requires in general two sets of anomalous dimensions $\alpha_\pm$~\cite{ddk}
\begin{equation}\label{gravad2} 
\alpha_\pm = -\frac{Q({\cal T})}{2} \pm \sqrt{\frac{Q^2({\cal T})}{4} + 
\frac{\varepsilon^2}{2}} 
\end{equation}
In Liouville theory it is 
common to ignore the $\alpha_-$ as leading to states that 
``do not exist'', as leading to non normalizable
states in the semiclassical limit where the 
central charge of the theory goes to infinity. 
This is what we have done so far.
However, in the context of string theory, with target-space time 
interpretation, the `wrong sign' states corresponding to $\alpha_-$ 
may not be excluded, and under 
certain conditions such ``wrong-sign'' dressing leads to physical states. 
This is our case here, since as we shall see below, one actually
does not face a situation with divergent central charge deficit 
which is cut off at a finite 
value at the ultraviolet world-sheet fixed point of the theory (Liouville scale $\phi_0 \to 0$). 
Using therefore opposite sign screening Liouville operators
for the two vertex operators corresponding to
the two $\Theta$ functions in (\ref{twothetas}), 
one encounters a supersymmetry breaking 
mass spectrum for the string theory at hand 
of the form: 
\begin{equation}
\Delta m_{\rm string}^2 \sim  2H{\rm cosh}\left(\frac{\varepsilon}{\sqrt{2}} \phi + 
\varepsilon X^0 \right)\Sigma_{45} 
\label{split2}
\end{equation}
It is evident that in such a case minimization of the potential energy 
in target space based on (\ref{split2}) will lead 
to the condition (\ref{liouvtime}), thereby
providing us with a {\it dynamical stability argument} in favour
of the identification of the Liouville world-sheet zero mode
with the target time. Physically, one may interpret this 
result as 
implying that a time-varying magnetic field of the form 
(\ref{twothetas}) induces 
back reaction of strings onto the space time 
in such a way that the mass splitiings of the string excitation
spectrum as a result of the field are actually stabilised.

From the Liouville action (\ref{liouvact}) we then observe that
in our case the dilaton field is 
\begin{equation}\label{dilaton}  
\Phi = Q\phi = Q^2 \varphi \sim (H^2 + u^2)  
\end{equation}
that is, one faces a situation with an asymptotically 
constant dilaton. 
This is a welcome fact, because otherwise, the space-time
would not be asymptotically flat, and one could face trouble in
appropriately defining masses~\footnote{For instance, in theories with
linear dilatons in time asymptotically~\cite{aben}, $q_0 X^0$, 
it is 
known that boson masses acquire tachyonic shifts $\delta m^2_B = q_0^2 $,
while fermion masses remain unaffected.}.

In the case of a constant dilaton the 
vacuum energy is determined by the central-charge charge 
deficit $Q^2$ (\ref{ratedeficit}), which in our case is: 
\begin{equation}
\Lambda = \frac{R^{2n}}{\phi_0^2}(H^2 + u^2)^2
\label{cosmoconst}
\end{equation}
where $\phi_0$ is the world-sheet zero mode of the rescaled
Liouville field (\ref{rescaling}).

It must be stressed that, due to the condition ${\rm Str}{\cal M}^2 =0$,
which is a characteristic feature of the magnetically-induced 
supersymmetry-breking  
scenario of \cite{bachas}, 
there are no quadratically divergent terms in the one-loop 
effective 
potential of the low-energy theory, which assumes the form~\cite{ferrara}:
\begin{eqnarray} \label{effpot}
&~& V_1 = V_0 + \frac{1}{64\pi^2}{\rm Str}{\cal M}^0 \Lambda _{uv}^4 {\rm ln}
\frac{\Lambda _{uv}^2}{\mu^2} + \frac{1}{32\pi^2} {\rm Str }{\cal M}^2 \Lambda_{uv}^2 
+ \frac{1}{64\pi^2}{\rm Str}{\cal M}^4 {\rm ln}\frac{{\cal M}^2}{\Lambda_{uv}^2} + \dots ~, \nonumber \\ 
&~& {\rm Str}{\cal M}^n = \sum_{i} (-1)^{2J_i}(2J_i - 1)m_i^n
\end{eqnarray}
where $\mu$ is a scale, and $V_0$ is a field-independent contribution. 
In our case $V_0$ is given by 
$\Lambda$ in (\ref{cosmoconst}). Note also that in a supersymmetric theory
(even if supersymmetry is broken) ${\rm Str}{\cal M}^0 =0$, due to a 
balance between fermionic and bosonic degrees of freedom.  
If the supersymmetry is broken at a TeV scale, then, the remaining 
${\rm Str}{\cal M}^4 {\rm ln}\frac{\cal M}{\Lambda_{uv}}$ term 
in (\ref{effpot}), 
which induces quadratic corrections to the 
Higgs mass, produces a stable hierarchy.

We now remark that the restoration of the conformal 
invariance by the Liouville mode results in the following equations
for the $\sigma$-model background fields/couplings 
$g^i$ near a fixed-point of the world-sheet renormalization group
(large-times cosmology) we restrict ourselves 
here~\cite{ddk,emn,schmid,dgmpp}:
\begin{equation}\label{liouvcond}
 (g^i)'' + Q (g^i)' = -\beta^i (g) 
\end{equation}
where the prime denotes derivative with respect to the 
Liouville zero mode $\phi_0$, and the sign 
on the right-hand-side is appropriate for supercritical 
strings~\cite{aben} we are dealing with here. 
In fact the $\beta^i$ functions satisfy a gradient flow property
\begin{equation}\label{flow}
\beta^j{\cal G}_{ij} = \frac{\delta C[g]}{\delta g^i}
\end{equation}
where ${\cal G}_{ij}=z^2 {\bar z}^2<V_i(z) V_j(0)>$ is the Zamolodchikov metric
in string theory space~\cite{zam}, with $V_i$ the appropriate vertex operators
corresponding to the couplings $g^i$, and $C[g]$ is the effective 
action which can be identified with the central charge deficit squared
$Q^2 [g,\phi_0]$ in our case.

It should be mentioned
for completeness
that the Liouville equations (\ref{liouvcond}), (\ref{flow}), 
which restore 
conformal invariance, can always be viewed as 
conformal invariance conditions of a $\sigma$-model in $d$+1 dimensional
space time, with the extra coordinate provided by the Liouville 
mode $\phi$. They themselves can be derived from a 
$d$+1 dimensional action, since the appropriate (Helmholtz) conditions
are statisfied in this case~\cite{emn}. 
Close to a fixed point, i.e. up to order $g^2$ in weak $\sigma$-model 
couplings/background fields, the action has the form~\cite{emn}:
\begin{equation}\label{fourdimaction} 
{\cal S} = \int d\phi_0 \left( \frac{1}{2}{g^i}' 
{\cal G}_{ij} {g^j}' - C[g]\right)
\end{equation} 
Indeed, it can be readily checked that the Lagrange equations in theory space
of this action reproduce the conformal invariance 
conditions (\ref{liouvcond}), 
provided ${\cal G}'_{ij} = Q {\cal G}_{ij}$, a property, which as 
explained in detail in \cite{emn}, characterizes Liouville dressing. 
Terms involving ${g^i}' \partial_m {\cal G}_{ij} {g^j}' $ are of 
order higher than $g^2$ and hence are ignored in our approach here. 
Notice that such an approach has also been used in \cite{schmid}
in discussing string cosmology and its relation to the renormalization-group
on the world sheet.

In our case $g^i$ is the metric $G_{\mu\nu}$, the dilaton $\Phi$ and 
the electromagnetic field $A_\mu$. 
The latter has already been discussed, and in our case,
for asymptotic times we are interested in, the dilaton is constant
(\ref{dilaton}). 
In what follows, therefore, we shall use (\ref{liouvcond}) to 
determine the form of the metric $G_{\mu\nu}$, 
assuming a Robertson-Walker
Universe with scale factor $a(t,\phi_0)$. 
We shall be interested only in the effective four-dimensional
low-energy theory, obtained by integrating our extra compact and bulk 
dimensions, as in \cite{dgmpp}. 

The relevant four-dimensional 
equations are (ignoring contributions from the recoil
velocities $u$ assumed of order lower than (or at most similar to) $H$): 
\begin{eqnarray}\label{metricliouv} 
&~& 3\frac{\ddot a}{a} = 0~, \nonumber \\
&~& -2\left(\frac{\dot a}{a}\right)^2 - \frac{\ddot a}{a}=
2\frac{a''}{a} + 2\left(\frac{a'}{a}\right)^2 + 
2\frac{H^2}{\phi_0}\frac{a'}{a}~, 
\end{eqnarray}
where the dot denotes derivative with respect to time $t=X^0$. 
From these equations we obtain the following solution for 
the scale factor (in string units):
\begin{equation}
a (t, \phi_0) = a_0 \phi_0^b~, \qquad b = \frac{1}{2} - \frac{H^2}{2}
\simeq \frac{1}{2}
\label{scalfinal}
\end{equation} 
We stress that this is the only acceptable solution from the 
point of view of Liouville dressing. The constant solution $b=0$,
which naively seems to be allowed, is excluded by the fact that 
such a solution corresponds to trivial gravitational dressing,
$g'=0$,  
which occurs {\it if and only if} 
$Q^2({\cal T}) =0$~\cite{ddk}
(critical-string, decoupling of the Liouville mode),
in contradiction to our case, where $Q^2 > 0$ (\ref{ratedeficit}).

We now recall 
(\ref{liouvtime}), according to 
which $\phi_0$ is related linearly to the 
cosmic time $X^0=-t$, $\phi_0 = \sqrt{2}t$. For asymptotically 
large times, therefore, this implies that 
the scale factor and the cosmological ``vacuum'' energy 
in our case behave as follows:
\begin{equation} 
a(t) \sim a_0 \sqrt{t}~, \qquad \Lambda \sim \frac{H^4R^{2n}}{t^2}  
\label{scalecosmoconst}
\end{equation}
We should remark that, since the dilaton is constant,  
the dilaton equation does not yield 
any further information apart from consistency checks,
which are easily performed. In 
particular, renormalizability of the $\sigma$-model
requires an additional constrain, namely the Curci-Paffutti 
equation~\cite{curci} which relates the dilaton $\beta$-function 
to the rest. This is valid for non-vanishing $\beta$-functions, 
and hence is applicable to our case as well~\cite{dgmpp}. It can be seen easily
that from this equation one obtains no other information than 
a consistency check 
on the scaling behaviour of the central charge deficit $Q^2$ 
obtained above (\ref{ratedeficit}).

From (\ref{scalecosmoconst}) we observe that for times $t$ 
of the order of the the Age of the observable Universe,  
$t  \sim 10^{60}$ in Planck units, and $H =10^{-30}$
as required by TeV scale supersymmetry breaking,
the cosmological vacuum energy is extremely suppressed at present
according to this model. Significantly 
larger relaxation rates are obtained if the recoil effects 
are the dominant ones, a case which will be discussed briefly below.
On the other hand, the $\sqrt{t}$ scaling of the scale factor
implies an asymptotically decelerating universe, ${\ddot a} \sim -t^{-3/2}$,
but on the other hand there is no cosmic horizon, and hence in this universe
one can define properly asymptotic states, and thus an S-matrix.

Notice, therefore, that in our non-critical string scenario,
one does indeed obtain an expanding Universe, in contrast to 
standard ekpyrotic scenaria~\cite{ekpyrotic,linde}, based on 
critical strings and specific solutions to classical equations 
of motion. Such scenaria correspond to a vanishing $Q^2$, and hence
$\beta^i=0$ as discussed above in (\ref{liouvcond}). In such a case,
the effective actions used in \cite{ekpyrotic,linde} are given by 
the flow function $C[g]$ for the specific set of backgrounds
used in those works. As we have seen, the presence of non-zero 
deficits $Q^2$ and  Liouville dependence leads to very different physics.

One of the most important features of the existence of a non-equilibrium 
phase of string theory due to the collision is the possibility
for an {\it inflationary phase}. Although the physics near the collision
is strongly coupled, and the $\sigma$-model perturbation theory
is not reliable, nevertheless one can give compelling 
physical arguments favouring the existence of an
early phase of the brane world where the four-dimensional 
Universe 
scale factor undergoes  exponential growth (inflation). 
This can be understood as follows: 
in our model we encounter two type-II string theory branes colliding, and then 
bouncing back. From a stringy point of view the collision and bounce 
will be described by a phase where open strings stretch between the 
two branes worlds (which can be thjought of as lying a few string scales
apart during the collision). During that ealy phase the excitation
energy of the brane worlds can be easily computed by the same methods
as those used to study scattering of type II D-branes in \cite{bachas2}.
Essentially, 
the time integral of 
the relevant potential energy 
yields the  
scattering amplitude for the two branes, which was 
computed in \cite{bachas2}. 
According to standard arguments of type II string theory 
the exchange of open strings between two parallel D-branes
is described by the emission of open-string pairs, and thus
an {\it annulus} world-sheet diagram~\footnote{This is in contrast to the 
type I string case where the corresponding exchange of open strings 
is described by 
a world-sheet disk to lowest order.}. As a 
result of the annulus graphs, the exchange of pairs of open strings
results in the appearance of ``spin structure factors'' 
in the scattering amplitude, which are expressed in terms of 
appropriate sums over Jacobi $\Theta$ functions. 
In particular, for small relative 
velocities $u \ll 1$ of the colliding branes, the appropriate 
spin structures start of at quartic order in $u$~\cite{bachas2}:
\begin{equation} 
\sum_{\alpha =2,3,4} e_\alpha \Theta_\alpha(u|\tau)\Theta_\alpha^3(0|\tau) \sim {\cal O}(u^4)~, \qquad e_2=-e_3=e_4=1
\end{equation}
This is a result of the property of the Jacobi functions 
that are even functions of theoir argument, as well as that the
$\Theta$ function satisfies by definition a ``diffusion'' equation:
$[\partial_\tau + \frac{i}{4\pi}\partial^2_\nu ]\Theta_\alpha (\nu|\tau)=0$.
The resulting excitation energy is therefore 
of order ${\cal O}(u^4)$ and may be thought 
off as an initial value of the central charge deficit of the 
non-critrical string theory describing the 
physics of our brane world after the collision. The deficit $Q^2$ 
is 
thus cut off at a finite value in the infrared, 
and hence one never encounters
a semiclassical limit for the underlying world-sheet field theory.
This justifies, as already mentioned, the use of ``wrong-sign'' Liouville
screening operators in this case. 
One may plausibly assume that the central charge deficit remains constant
for some time, which is the era of {\it inflation}, as expressed 
by the equation (\ref{liouvcond}) for the scale factor. For (finite) 
constant 
$Q^2=Q_*^2= {\cal O}(u^4)$ it is easy to infer from (\ref{liouvcond}) 
a scale factor exponentially growing with the Liouville zero mode  
$a(\phi_0) = e^{Q_*\phi_0/2}$. Upon the condition (\ref{liouvtime}), then, 
one obtains an ealy inflationary phase after the collision, in contrast 
to 
the critical-string based arguments of \cite{linde}. The duration of 
the inflationary phase is $t_{\rm inf} \sim 1/Q_* \sim {\cal O}(u^{-2})$,
which yields the conventional values of inflationary models
of order $t_{\rm infl} \sim 10^9 t_{\rm Planck}$ for $u^2 \sim 
10^{-9}$. This is compatible with the non-relativistic
approximation for the D-branes, where our formalism is valid.
Note that for such values of $u$ the recoil effect is the dominant one in 
the relaxation of the vacuum energy (\ref{cosmoconst}), while the 
magnetic field is mainly responsible only for the supersymmetry breaking.

A final comment on the issue of inflation concerns the role of the dilaton
as an inflaton field. During the phase of constant $Q_*$ one may imagine 
the appearance of a  scalar dilaton field in $(X^0,{\vec x})$ space time
which is linear in $X^0$: $\Phi = Q_* X^0$ such that after the condition
(\ref{liouvtime}) it cancels any dilaton effects, in the sense of a 
trivial world-sheet curvature coupling. 
This is a consistent solution of the conditions (\ref{liouvcond}),
implying a constant dilaton $\beta$-function $Q_*^2$, as required
in non-critical strings with constant central-charge  
deficit~\cite{aben}. 
Note that in this scenario, asymptotically in time,
the dilaton $\Phi(X^0,{\vec x})$ tends to a constant, so {\it on the
hypersurface} (\ref{liouvtime}) of the D+1 extended space time
resulting after Liouville dressing 
the dilaton plays no actual r\^ole in the 
scenario. However, we stress that 
there is a dilaton field
$\Phi(X^0,{\vec x})$   
at the initial stage after the collision, 
which is non trivial 
on the D+1 dimensional extended space time,
and hence one can safely speak about a dilaton acting like an inflaton field 
in this scenario.

Before closing we would like to make some brief
remarks on the possible relevance of this toy model
to realistic present-era cosmology, although we stress again,
this is not the main purpose of our work. 
If one takes into account 
recent astrophysical claims~\cite{carroll} 
according to which 
the present era of the Unvierse appears accelerating, 
then our results above seem to be ruled out by experiment. 
Of course the naive way out would be to observe 
that the above results
are valid for times 
much later than the present era where one sees the acceleration. 

Another interesting feature is the 
order of magnitude of the present-era vacuum energy.  
Physically, the relaxation to zero of the vacuum energy we find here 
seems quite plausible, given the transient nature of the collision
of the two branes. It is interesting to notice 
that its order of magnitude depends crucially
on which is the dominant effect in the relaxation rate (\ref{cosmoconst}).
If one insists on getting an inflationary era that lasts according to 
arguments of standard
cosmology, then, as we have seen above,
one requires recoil velocities of order $u < 10^{-4}$, which 
imply that recoil of the colliding branes is the dominant effect 
in providing (long after the scattering event) a vacuum-energy 
relaxation rate of order $u^4/t^2$ upon the identification (\ref{liouvtime})
of Liouville mode with target time. This 
would then yield a vacuum energy which lies comfortably within the current 
observations~\cite{timedep,dgmpp}~\footnote{Notice 
that the coefficient of $1/t^2$
is of the same order as the initial vacuum energy $Q_*^2 \sim u^4$ during the 
inflationary era for type II strings.  
This suggests that a natural interpolating function
for $Q^2(t)$ from the end of inflation $t_0 \sim 10^9$ (in Planck (string) 
units) until the present era
would be $Q^2(t) = u^4/[(t-t_0)^2 + {\cal O}(1)]$. At present, however, 
we cannot support this by any quantitative analysis.}.  

In this latter respect, however, 
an interesting observation can be made regarding our results. 
Notice that in case one does not care too much 
about the order of magntidude of the duration of   
the inflationary phase, but restricts oneself to the case
where the recoil of the branes is subdominant as compared
to the magnetic field effects, then the coefficient
of the $1/\phi_0^2$ scaling in (\ref{cosmoconst}) 
is of order $H^4$, which by itself
yields the order $10^{-120}$ in Planck units, 
in the case of supersymmetry breaking advocated here.
This value is of the same order as the one 
claimed by the astrophysicists
to have been ``observed'' from the preliminary superonvae data
for the current era cosmological
vacuum energy, 
Interestingly enough, therefore, the order of this 
coefficient by itself is what one needs~\cite{banks}
to resolve the supersymmetry-breaking/cosmological constant hierarchy.
In our case naively, one could have obtained this latter result 
had one not made the connection of the Liouville scale
with the time (\ref{liouvtime}). Indeed, in such a case, where 
$\phi$ and $X^0$ are independent variables,
which notably is mathematically consistent,
one can freeze the renormalization group scale $\phi_0$ 
to order $R^n$, to obtain a vacuum energy contribution 
(\ref{cosmoconst}) of the 
required magnitude. This  
vacuum energy is independent of time, and the 
solution of the metric equations (\ref{metricliouv}) 
varies only with respect to the scale $\phi_0$ (c.f (\ref{scalfinal})). 
Notice that this scenario is compatible with the alternative 
way of parametrizing the adiabatic switching  on of the 
magnetic field on the observable brane, $H(1-e^{-\varepsilon X^0})$. 

However, as we have explained above, the transient 
nature of the colliding branes scenario, we are advocating here, 
seems to imply that the correct physical
picture is the one in which (\ref{liouvtime}) is valid and the excitation
energy of the non-equilibrium system 
is given by a relaxing-to-zero 
time-dependent $\Lambda$ (\ref{scalecosmoconst}), 
as in quintessence models~\cite{carroll}.
The equilibrium state, which is the true ground state of the relaxing 
system, is then only reached asymptotically in time, 
and in our case has vanishing energy, 
due to the cancellation of the (positive) supersymmetry
breaking energy contribution $H^2$ by the opposite in sign vacuum 
energy ocontribution of the 
negative 
tension brane, assumed to be our world.
As we mentioned above, the initial instability 
due to the presence of negative tension branes is not necessarily
a drawback in our cosmological non equilibrium framework. 
The phenomenology of this transient scenario, therefore, seems to favour
the recoil effect as the dominant one in the vacuum energy relaxation rate,
without affecting our previous arguments on supersymmetry breaking
which solely occurs due to the magnetic field.

This concludes our discussion on this toy model.
It would be interesting to attempt and construct   
phenomenologically realistic supersymmetric brane-Universe 
models along the lines
outlined above, exhibiting an  
accelerating phase at late times. Such a situation is encountered 
in the non-supersymmetric
case of \cite{dgmpp}. The hope is that a realistic stringy 
Universe model can be found, 
which  has a late times accelerating phase
and is capable of resolving 
the hierarchy between the supersymmetry-breaking scale and the 
present-era 
cosmological vacuum energy. This is left for future work.

\section*{Acknowledgements}

The work of E.G. is supported 
by a King's College London Research Studentship (KRS).
This work is also partially supported by the European Union
(contract ref. HPRN-CT-2000-00152).

\end{document}